\documentclass{PNASTWO}

\url{www.pnas.org/cgi/doi/10.1073/pnas.1418994112}
\copyrightyear{2015}
\issuedate{Issue Date}
\volume{Volume}
\issuenumber{Issue Number}
\begin{document}

\title{Suppression of phase separation and giant enhancement of superconducting transition temperature in FeSe$_{1-x}$Te$_{x}$ thin films}

\author{Yoshinori Imai\affil{1}{Department of Basic Science, the University of Tokyo, Tokyo, Japan},
Yuichi Sawada\affil{1}{},
Fuyuki Nabeshima\affil{1}{},
\and
Atsutaka Maeda\affil{1}{}}

\contributor{Submitted to Proceedings of the National Academy of Sciences
of the United States of America}

%%%Newly updated.
%%% If significance statement need, then can use the below command otherwise just delete it.
\significancetext{
To clarify the mechanism of superconductivity in iron-based superconductors, it is crucial to investigate FeSe$_{1-x}$Te$_x$, which has the simplest crystal structure among them.
There is, however, a serious obstacle to the understanding of its superconductivity; phase separation occurs in the region of $0.1 \le x \le 0.4$, and thus a whole phase diagram has not been available.
%One of the biggest obstacles to the understanding of its superconductivity is that phase separation occurs in the region of $0.1 \le x \le 0.4$, and that a perfect phase diagram is not available.
Here we report the successful fabrication of FeSe$_{1-x}$Te$_x$ films with $0 \le x \le 1$.
This is the first demonstration of the suppression of the phase separation.
What is more notable is that a giant enhancement of $T_\mathrm{c}$ is observed in the ``phase-separation region''.
%, and that a sudden suppression of $T_\mathrm{c}$ occurs at $x=0.1-0.2$.
The complete phase diagram that we present provides a novel perspective for the mechanism of superconductivity of this material.
}

\maketitle

\begin{article}
\begin{abstract}
{We demonstrate the successful fabrication on CaF$_2$ substrates of FeSe$_{1-x}$Te$_{x}$ films with $0 \le x \le 1$, including the region of $0.1 \le x \le 0.4$, which is well known to be the ``phase-separation region'', via pulsed laser deposition which is a thermodynamically non-equilibrium method.
In the resulting films, we observe a giant enhancement of the superconducting transition temperature, $T_\mathrm{c}$, in the region of $0.1 \le x \le 0.4$: the maximum value reaches 23 K, which is approximately 1.5 times as large as the values reported for bulk samples of FeSe$_{1-x}$Te$_{x}$.
We present a complete phase diagram of FeSe$_{1-x}$Te$_{x}$ films.
Surprisingly, a sudden suppression of $T_\mathrm{c}$ is observed at $0.1<x<0.2$, while $T_\mathrm{c}$ increases with decreasing $x$ for $0.2 \le x < 1$.
Namely, there is a clear difference between superconductivity realized in $x=0-0.1$ and in $x \ge 0.2$.
To obtain a film of FeSe$_{1-x}$Te$_{x}$ with high $T_\mathrm{c}$, the controls of the Te content $x$ and the in-plane lattice strain are found to be key factors.}
\end{abstract}


\abbreviations{PLD, pulsed laser deposition; XRD, X-ray diffraction}

\dropcap{S}ince the discovery of superconductivity in LaFeAs(O,F)\cite{Kamihara08}, many studies concerning iron-based superconductors have been conducted.
FeSe is the iron-based superconductor with the simplest crystal structure\cite{Wu08}.
The $T_\mathrm{c}$ of FeSe is approximately 8 K, which is not very high in comparison with other iron-based superconductors.
However, the value of $T_\mathrm{c}$ strongly depends on the applied pressure, and the temperature at which the resistivity becomes zero, $T_\mathrm{c}^\mathrm{zero}$, reaches as high as $\sim 30$ K at 6 GPa\cite{PhysRevB.81.205119}.
This suggests that FeSe samples with higher $T_\mathrm{c}$ are available by the fabrication of thin films because we can introduce lattice strain.
Indeed, we have previously reported that FeSe films fabricated on CaF$_2$ substrates exhibit $T_\mathrm{c}$ values approximately 1.5 times higher than those of bulk samples because of in-plane compressive strain\cite{Nabe13}.
On the other hand, superconductivity with $T_\mathrm{c}$ of 65 K has recently been reported in a monolayer FeSe film on SrTiO$_3$\cite{CPL.29.37402,NatCommun.12.605}.
It is unclear whether this superconductivity results from the characteristics of the interface.
However, this finding indicates that FeSe demonstrates potential as a very-high-$T_\mathrm{c}$ superconductor.

The partial substitution of Te for Se in FeSe also raises $T_\mathrm{c}$ to a maximum of 14 K at $x=0.5-0.6$\cite{Fang08}.
In FeSe$_{1-x}$Te$_x$, it is well known that we cannot obtain single-phase samples with $0.1 < x < 0.4$ because of phase separation\cite{Fang08}.
Here, we focus on this region of phase separation.
Generally, the process of film deposition involves crystal growth in a thermodynamically non-equilibrium state.
Thus, film deposition provides an avenue for the synthesis of a material with a metastable phase. 
In this letter, we report the fabrication of epitaxial thin films of FeSe$_{1-x}$Te$_x$ with $0 \le x \le 1$ on CaF$_2$ substrates using the pulsed laser deposition (PLD) method.
We demonstrate that single-phase epitaxial films of FeSe$_{1-x}$Te$_x$ with $0.1 \le x \le 0.4$ are successfully obtained and that the maximum value of $T_\mathrm{c}$ is as large as 23 K, which is higher than the previously-reported values for bulk and film samples of FeSe$_{1-x}$Te$_x$\cite{Fang08,Bellingeri10,Tsukada11,APL.99.202503,NatCommun.4.1347, APL.104.262601}, except for those of the monolayer FeSe films\cite{CPL.29.37402,NatCommun.12.605}.
Our results clearly show that the optimal Te content for the highest $T_\mathrm{c}$ for FeSe$_{1-x}$Te$_x$ films on CaF$_2$ is different from the widely-believed value for this system.

\begin{figure*}[t]
%h=here, t=top, b=bottom, p=separate figure page
\begin{center}%\leavevmode
\includegraphics[width=1\linewidth]{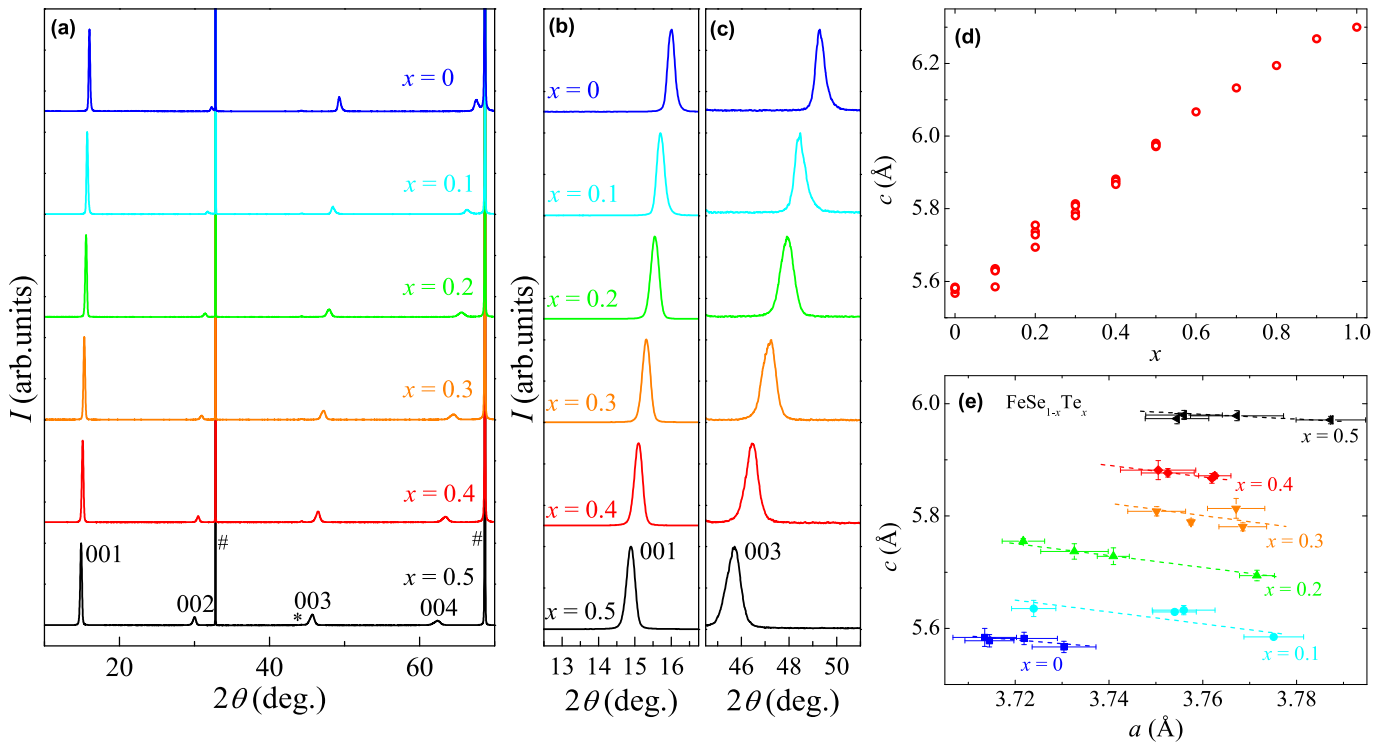}
%\vspace{10mm}
\caption{
(a) Out-of-plane X-ray diffraction patterns of FeSe$_{1-x}$Te$_x$ thin films for $x=0-0.5$  with film thicknesses of 120-147 nm. 
The number signs represent peaks associated with the substrate.
The asterisk represents an unidentified peak.
Enlarged segments of the plots presented in (a) near the 001 and 003 peaks are shown in (b) and (c), respectively.
(d) The $c$-axis lengths of FeSe$_{1-x}$Te$_x$ films, where the $x$ value indicated on the horizontal axis is the nominal Te content.
(e) Relations between the $a$-axis and $c$-axis lengths in FeSe$_{1-x}$Te$_x$ films.
The colors and shapes of the symbols correspond to the Te content $x$, as shown in the figure.
The dashed lines are guides for the eye.
The data for $x=0$ and $0.5$ presented in (a)-(e) are cited from our previous papers\cite{Nabe13,Maeda14}.
}
\label{fig:xrd}
\end{center}
%\vspace{-5.5mm}
\end{figure*}
%
%\section{Experimental}

%\section{Results and discussion}
    Figure \ref{fig:xrd}a presents the X-ray diffraction patterns of FeSe$_{1-x}$Te$_{x}$ films for $x=0-0.5$ on CaF$_2$.   
    Here and hereafter, the Te content $x$ of our films represents the nominal Te composition of the polycrystalline target.
    With the exception of an unidentified peak in the FeSe$_{0.5}$Te$_{0.5}$ film, only the $00l$ reflections of a tetragonal PbO-type structure are observed, which indicates that these films are well oriented along the $c$ axis.
    Figures \ref{fig:xrd}b and \ref{fig:xrd}c present enlarged segments of these plots near the 001 and 003 reflections, respectively.
    The 2$\theta$ values of the peak positions decrease with increasing $x$ in a continuous manner, which is consistent with the fact that the $c$-axis length increases with increasing $x$.
    It should be noted that the values of the full widths at half maximum (FWHM), $\delta(2\theta)$, of the FeSe$_{1-x}$Te$_{x}$ films with $x=0.1-0.4$, which is known as the region of phase separation in the bulk samples\cite{Fang08}, are $\delta(2\theta) = 0.2-0.3^\circ$ for the 001 reflection and $\delta(2\theta) = 0.4-0.6^\circ$ for the 003 reflection, which are nearly the same as the values for the FeSe and FeSe$_{0.5}$Te$_{0.5}$ films.
    This result is in sharp contrast to the previously-reported result that the FWHM was broad $only$ in films of FeSe$_{1-x}$Te$_{x}$ with $x=0.1$ and $0.3$,\cite{Wu09} where phase separation has been believed to occur.
    The results presented in Figs. \ref{fig:xrd}a-\ref{fig:xrd}c indicate the formation of a single phase in our FeSe$_{1-x}$Te$_{x}$ films with $x=0.1-0.4$.

    In Figure \ref{fig:xrd}d, the $c$-axis lengths of 29 films of FeSe$_{1-x}$Te$_{x}$ are plotted as a function of $x$.
    The values of the $c$-axis lengths vary almost linearly with the nominal Te contents of the targets in the whole range of $x$ including both end-member materials.
    The evident formation of a single phase and the systematic change in the $c$-axis length strongly indicate that the nominal Te content of the polycrystalline target is nearly identical to that of the final FeSe$_{1-x}$Te$_{x}$ film.
    Note that the compositional analysis of grown films using scanning electron microscopy/energy-dispersive X-ray (SEM/EDX) analysis is impossible for FeSe$_{1-x}$Te$_{x}$ films on CaF$_2$ substrates because the energies of the K-edge of Ca and the L-edge of Te are very close to each other.
    The above-mentioned features indicate that phase separation is suppressed in our FeSe$_{1-x}$Te$_{x}$ films with $x=0.1-0.4$ on CaF$_2$ substrates.
    To our knowledge, this result is the first manifestation of the suppression of phase separation in FeSe$_{1-x}$Te$_{x}$ with $x=0.1-0.4$.

    Figure \ref{fig:xrd}e presents the relations between the $a$-axis and $c$-axis lengths in films of FeSe$_{1-x}$Te$_{x}$.
    At first glance, there seems to be no relations between the $a$-axis length and $x$, in sharp contrast to the behavior of the $c$-axis length.
    The $a$-axis and $c$-axis lengths of films with the same $x$ show a weak negative correlation. 
    This behavior cannot be explained by a difference in Te content of a film, which should result in a positive correlation. 
    By contrast, if variations in $c$ are caused by a difference in in-plane lattice strain, this behavior can be explained in terms of the Poisson effect. 
    Indeed, the $a$-axis lengths of films of FeSe and FeSe$_{0.5}$Te$_{0.5}$ are smaller than those of bulk samples with the same composition.
    Thus, we consider that the $a$-axis length predominantly depends on the in-plane lattice strain rather than the Te content $x$.
%    The $a$-axis lengths of films of FeSe and FeSe$_{0.5}$Te$_{0.5}$ are smaller than those of bulk samples with the same composition, which shows our films experience a compressive strain.
    One might think that this behavior looks strange, because the lattice constant of CaF$_2$($a_\mathrm{CaF_2}/\sqrt{2}$) is longer than the $a$ of FeSe$_{1-x}$Te$_{x}$, which usually leads to a tensile strain.
    In the previous paper, the penetration of F$^-$ ions from the CaF$_2$ substrates into the films has been proposed as a possible mechanism for nontrivial compressive strain in  FeSe$_{1-x}$Te$_{x}$ films on CaF$_2$ substrates\cite{sust.26.075002}.
    Because of smaller ionic radius of F$^-$ than that of Se$^{2-}$, this peculiar compressive strain can be explained by the partial substitution of F$^-$ for Se$^{2-}$ near the interface between a film and a substrate. 
%    Additionally, the $a$-axis lengths do not show a monotonic change with increasing film thickness, while an usual compressive strain resulted from the lattice mismatch between a substrate and an overlayer will be relaxed with increasing film thickness(see the supporting information).

\begin{figure*}[t]
%h=here, t=top, b=bottom, p=separate figure page
\begin{center}%\leavevmode
\includegraphics[width=0.9\linewidth]{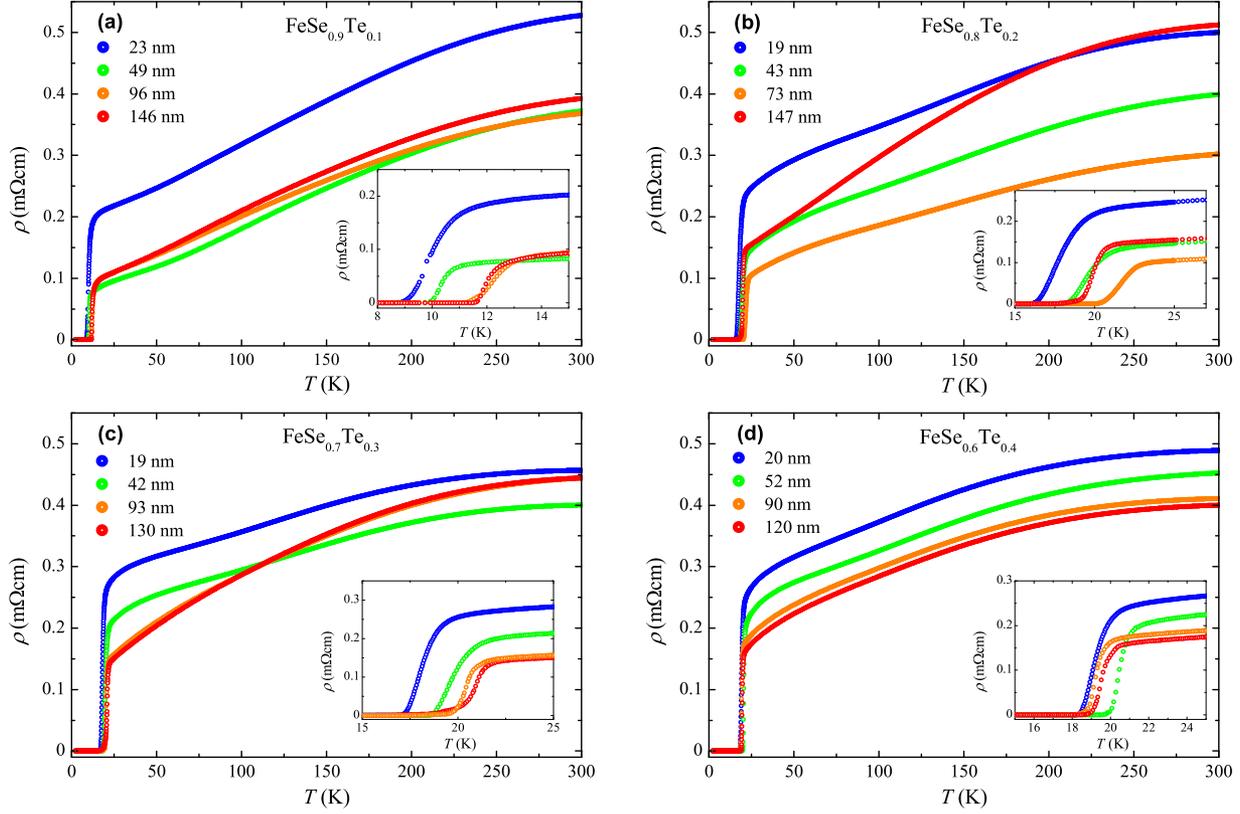}
%\vspace{10mm}
\caption{
Temperature dependences of the electrical resistivities, $\rho$, of FeSe$_{1-x}$Te$_x$ thin films for (a) $x=0.1$, (b) $x=0.2$, (c) $x=0.3$ and (d) $x=0.4$ with different film thicknesses.
The insets present enlarged views of the plots near the superconducting transition.
}
\label{fig:rho}
\end{center}
%\vspace{-5.5mm}
\end{figure*}

    Figures \ref{fig:rho}a-\ref{fig:rho}d present the temperature dependences of the electrical resistivities, $\rho$, of 16 films of FeSe$_{1-x}$Te$_{x}$ for $x=0.1-0.4$. 
    The value of $T_\mathrm{c}$ depends on the film thickness, even in films with the same $x$.
    The highest $T_\mathrm{c}^\mathrm{onset}$, which is defined as the temperature where the electrical resistivity deviates from the normal-state behavior, and the $T_\mathrm{c}^\mathrm{zero}$ of the FeSe$_{1-x}$Te$_{x}$ films are 13.2 K and 11.5 K, respectively, for $x=0.1$; 22.8 K and 20.5 K, respectively, for $x=0.2$; 20.9 K and 19.9 K, respectively, for $x=0.3$; and 20.9 K and 20.0 K, respectively, for $x=0.4$.
    Compared with the results for bulk samples, a drastic enhancement of $T_\mathrm{c}$ is observed in these FeSe$_{1-x}$Te$_{x}$ films.
    Surprisingly, the values of $T_\mathrm{c}^\mathrm{zero}$ in the films with $x=0.2$ and $0.4$ exceed 20 K.  
    These values are larger than those reported for FeSe$_{0.5}$Te$_{0.5}$ films\cite{Maeda14,Bellingeri10,APL.99.202503,NatCommun.4.1347}.
    In particular, the $T_\mathrm{c}$ of the FeSe$_{0.8}$Te$_{0.2}$ film with a thickness of 73 nm is approximately 1.5 times as high as those of bulk crystals of FeSe$_{1-x}$Te$_{x}$ with the optimal composition, $x \approx 0.5$\cite{Fang08}.
    Based on the measurement of the $\rho$ of the FeSe$_{0.8}$Te$_{0.2}$ film under a magnetic field applied along the $c$ axis, we estimate an upper critical field at 0 K of $\mu_0 H_\mathrm{c2}=55.4$ T using the Werthamer-Helfand-Hohenberg(WHH) theory\cite{whh}, which yields a Ginzburg-Landau coherence length at 0 K of $\xi_{ab} (0) \sim 24.4$ \AA  (see the supporting information).
    This value of $\mu_0 H_\mathrm{c2}$ is approximately half the value for an FeSe$_{0.5}$Te$_{0.5}$ film on CaF$_2$ with a $T_\mathrm{c}$ of approximately 16 K.\cite{Tsukada11}

    Using the data shown above, we present the phase diagram of FeSe$_{1-x}$Te$_{x}$ films on CaF$_2$ substrates in Fig. 3.
    For comparison, the data for bulk samples of FeSe$_{1-x}$Te$_{x}$\cite{Fang08, JPSJ.79.084711} are also plotted in this figure.
    In bulk crystals, the optimal Te content to achieve the highest $T_\mathrm{c}$ is considered to be $x \approx 0.5$, and phase separation occurs in the region of $0.1 \le x \le 0.4$\cite{Fang08}.
    However, our data clearly demonstrate that this phase separation is absent and that the optimal composition for an FeSe$_{1-x}$Te$_{x}$ film on a CaF$_2$ substrate is not $x \approx 0.5$ but $x \approx 0.2$.    
    It should be noted that the dependence of $T_\mathrm{c}$ on $x$  suddenly changes at the boundary defined by  $0.1 < x < 0.2$.
    Unlike the ``dome-shaped'' phase diagram that is familiar in iron-based superconductors, the values of $T_\mathrm{c}$ in films with $0.2 \le x \le 1$ increase with decreasing $x$, while the strong suppression of $T_\mathrm{c}$ is observed at $0.1<x<0.2$.
    The behavior in films with $x \ge 0.2$ can be explained by the empirical law which shows the relation between $T_\mathrm{c}$ and structural parameters.
    In iron-based superconductors, it is well accepted that the bond angle of ($Pn$, $Ch$)-Fe-($Pn$, $Ch$) ($Pn=Pnictogen$, $Ch=Chalcogen$), $\alpha$,\cite{LeePlot, LeePlot2} and/or the anion height from the iron plane, $h$,\cite{AnionHeight} are the critical structural parameters that determine the value of $T_\mathrm{c}$.
    In bulk samples of FeSe$_{1-x}$Te$_{x}$, $\alpha$ and $h$ approach their optimal values, i.e. $\alpha = 109.47 ^\circ$\cite{LeePlot, LeePlot2} and $h=1.38$ \AA\cite{AnionHeight}, with decreasing $x$ (down to $x=0$), which should be the same in FeSe$_{1-x}$Te$_{x}$ films.
    Therefore, the increase of $T_\mathrm{c}$ in films with $0.2 \le x \le 1$ with decreasing $x$ can be explained by the optimization of $\alpha$ and/or $h$ based on the empirical law.
    However, the sudden suppression of $T_\mathrm{c}$ in films with $0 \le x < 0.2$ is not consistent with this scenario, and its origin should be sought among other factors. 
    We consider there are two candidates for this origin from the structural analysis of bulk samples of FeSe$_{1-x}$Te$_{x}$.
    One is the effect of the orthorhombic distortion.
    In a bulk sample of FeSe, a structural phase transition from tetragonal to orthorhombic occurs at 90 K\cite{PhysRevLett.103.057002}.
    However, in bulk samples of FeSe$_{1-x}$Te$_{x}$ with $x \sim 0.4-0.6$ which $T_\mathrm{c}$s take optimum values, there are papers with different conclusions on the presence/absence of the similar type of the structural transition as that of FeSe.\cite{JPSJ.78.074718,PhysRevB.79.054503,JACS.131.16944}
    It should be noted that a structural transition temperature is lower and that the orthorhombicity is much smaller than those of FeSe even in the report where the structural transition is present.\cite{JACS.131.16944}
    These results on crystal structures suggest that the orthorhombic distortion results in a suppression of $T_\mathrm{c}$.
    This scenario is applicable to the behavior of our films, if a large orthorhombic distortion is observed only in films with $x=0-0.1$.
    The other candidate is the change in the distance between the layers of Fe-$Ch$ tetrahedra, $\delta$.
     As shown in the supporting information, in polycrystalline samples of FeSe$_{1-x}$Te$_{x}$, the $\delta$ value of FeSe is much smaller than those of FeSe$_{1-x}$Te$_{x}$ with $x \ge 0.5$ at which $\delta$ is nearly independent of $x$\cite{JPSJ.78.074718}.
     We speculate that the decrease of $\delta$  in FeSe is related with the suppression of $T_\mathrm{c}$.
    Indeed, in polycrystalline samples, FeSe exhibits smaller values of $\delta$ and $T_\mathrm{c}$ than does FeSe$_{0.5}$Te$_{0.5}$\cite{Fang08, JPSJ.78.074718}, and the intercalation of alkali metals and alkaline earths into FeSe result in the $c$-axis length as large as approximately 20 \AA \ and  $T_\mathrm{c}$ as high as 45 K.\cite{SciRep.2.426, JPSJ.82.123705}
    At this moment, the origin of the suppression of $T_\mathrm{c}$ at $0.1<x<0.2$ is unclear, which will be a subject of a future study.
    Regardless of its origin, we believe that it is reasonable to distinguish between superconductivity in $x=0-0.1$ and in $x \ge 0.2$.
    In other words, our phase diagram of Fig. 3 provides a new view for superconductivity in FeSe$_{1-x}$Te$_{x}$, that is, a discontinuity in superconductivity of FeSe$_{1-x}$Te$_{x}$.
    We are able to come to this picture, only after the data for $x=0.1-0.4$ become available in this study.
    If we remove a cause for the suppression of $T_\mathrm{c}$ in $x=0, 0.1$ in some way, a further increase in $T_\mathrm{c}$ can be expected because of the optimization of structural parameters.

\begin{figure}[t]
%h=here, t=top, b=bottom, p=separate figure page
\begin{center}%\leavevmode
\includegraphics[width=0.99 \linewidth]{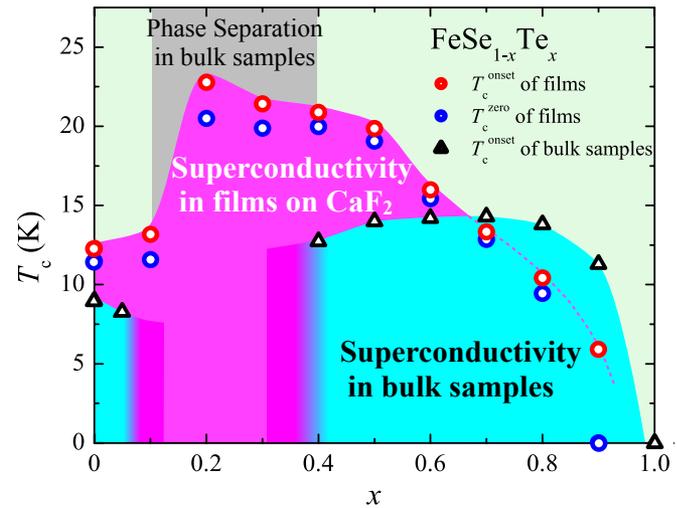}
%\vspace{-5mm}
\caption{
Dependence of $T_\mathrm{c}$ on $x$.  
The red and blue circles represent the $T_\mathrm{c}^\mathrm{onset}$ and $T_\mathrm{c}^\mathrm{zero}$ values of the FeSe$_{1-x}$Te$_x$ thin films, respectively.  
The black triangles represent the $T_\mathrm{c}^\mathrm{onset}$ values obtained in measurements of the magnetic susceptibility of bulk samples\cite{Fang08,JPSJ.79.084711}.
The dashed curve is a guide for the eye.
}
\label{fig:sozu}
\end{center}
%\vspace{5.5mm}
\end{figure}

    In conclusion, we prepared high-quality epitaxial thin films of FeSe$_{1-x}$Te$_{x}$ on CaF$_2$ substrates using the pulsed laser deposition method.
    We successfully obtained FeSe$_{1-x}$Te$_{x}$ films with $0.1 \le x \le 0.4$, which has long considered to be the ``phase-separation region'', using a thermodynamically non-equilibrium growth of film deposition.
    From the results of electrical resistivity measurements, a complete phase diagram is presented in this system, in which the maximum value of $T_\mathrm{c}$ is as high as 23 K at $x=0.2$.
    Surprisingly, a sudden suppression of $T_\mathrm{c}$ is observed at $0.1<x<0.2$, while $T_\mathrm{c}$ increases with decreasing $x$ for $0.2 \le x < 1$.
    This behavior is different from the ``dome-shaped'' phase diagram that is familiar in iron-based superconductors.
%    The electrical resistivities of FeSe$_{1-x}$Te$_{x}$ films with $x=0.2$ and $0.4$ became zero above 20 K. 
%    In order to obtain a film of FeSe$_{1-x}$Te$_{x}$ with high $T_\mathrm{c}$, the controls of the Te content $x$ and the in-plane lattice strain are found to be key factors.
\\

\begin{materials}
%\section{Film Fabrication} 
All of the FeSe$_{1-x}$Te$_{x}$($x=0-1$) films in this study were grown by the pulsed laser deposition method using a KrF laser\cite{Imai09, Imai10}.  
Polycrystalline pellets with a nominal composition of FeSe$_{1-x}$Te$_{x}$($x=0-1$)  were used as the targets.
The substrate temperature, repetition rate, and back pressure were 280 $^{\circ}$C, 20 Hz, and 10$^{-7}$ Torr, respectively. 
Single crystals of the CaF$_2$ (100), which is one of the most preferred materials for the thin-film growth of FeSe$_{1-x}$Te$_{x}$\cite{Tsukada11, Nabe13}, were used as the substrates.
We used a metal mask to prepare the FeSe$_{1-x}$Te$_{x}$ films in a six-terminal shape for transport measurements.  
The thicknesses of the thin films were measured using a Dektak 6 M stylus profiler and estimated to be 12-148 nm. 
The crystal structures and orientations of the films were characterized by four-circle X-ray diffraction (XRD) with Cu K$\alpha$ radiation at room temperature.
The $a$-axis and $c$-axis lengths are determined from the $204$ and $00l$ reflections in XRD measurements, respectively.
Electrical-resistivity measurements were conducted using the four-terminal method from 2 K to 300 K with magnetic fields up to 9 T applied perpendicular to the film surface.
%\section{Characterization}
\end{materials}

\begin{acknowledgments}
We are grateful to Dr. Ichiro Tsukada at CRIEPI for fruitful discussions. 
The authors thank Profs. Jun-ichi Shimoyama and Kohji Kishio at Department of Applied Chemistry, the University of Tokyo for their especial support in the chemical-composition analysis of the films.

This work was partially supported by Strategic International Collaborative Research Program (SICORP), Japan Science and Technology Agency.
\end{acknowledgments}

%\bibliographystyle{pnas}
%\bibliography{database}

\providecommand{\noopsort}[1]{}\providecommand{\singleletter}[1]{#1}

\end{article}

\end{document}